\documentclass[12pt]{article}
\usepackage{graphicx}

      \usepackage[cp1251]{inputenc} %% Windows
      \usepackage[russian]{babel}   %%

\begin{document}
 \noindent {\footnotesize\it Astronomy Letters, 2013, Vol. 39, No. 8, pp. 532–549.}
\newcommand{\dif}{\textrm{d}}

 \noindent
 \begin{tabular}{llllllllllllllllllllllllllllllllllllllllllllll}
 & & & & & & & & & & & & & & & & & & & & & & & & & & & & & & & & & & & & & \\\hline\hline
 \end{tabular}

 \vskip 1.0cm

 \centerline{\bf Galactic Kinematics from a Sample of Young Massive Stars}
 \bigskip
 \centerline{\bf V.V. Bobylev$^{1,2},$ and A.T. Bajkova$^1$}
 \bigskip
 \centerline{\small $^{1}$\it Pulkovo Astronomical Observatory, St. Petersburg,  Russia}
 \centerline{\small $^{2}$\it Sobolev Astronomical Institute, St. Petersburg State University, Russia}
 \bigskip
 \bigskip

{\bf Abstract}—Based on published sources, we have created a
kinematic database on 220 massive $(>10M_\odot)$ young Galactic
star systems located within $\leq3$~kpc of the Sun. Out of them,
$\approx$100 objects are spectroscopic binary and multiple star
systems whose components are massive OB stars; the remaining
objects are massive Hipparcos B stars with parallax errors of no
more than 10\%. Based on the entire sample, we have constructed
the Galactic rotation curve, determined the circular rotation
velocity of the solar neighborhood around the Galactic center at
$R_0=8$~kpc, $V_0=259\pm16$~km s$^{-1},$ and obtained the
following spiral density wave parameters: the amplitudes of the
radial and azimuthal velocity perturbations $f_R=-10.8\pm1.2$~km
s$^{-1},$ and $f_\theta=7.9\pm1.3$~km s$^{-1},$ respectively; the
pitch angle for a two-armed spiral pattern
$i=-6.0^\circ\pm0.4^\circ$, with the wavelength of the spiral
density wave near the Sun being $\lambda=2.6\pm0.2$~kpc; and the
radial phase of the Sun in the spiral density wave
$\chi_\odot=-120^\circ\pm4^\circ$. We show that such peculiarities
of the Gould Belt as the local expansion of the system, the
velocity ellipsoid vertex deviation, and the significant
additional rotation can be explained in terms of the density wave
theory. All these effects decrease noticeably once the influence
of the spiral density wave on the velocities of nearby stars has
been taken into account. The influence of Gould Belt stars on the
Galactic parameter estimates has also been revealed. Eliminating
them from the kinematic equations has led to the following new
values of the spiral density wave parameters:
$f_\theta=2.9\pm2.1$~km s$^{-1}$ and
$\chi_\odot=-104^\circ\pm6^\circ$.

%DOI: 10.1134/S106377371308001X

\newpage
\section*{INTRODUCTION}
A wide variety of data are used to study the Galactic kinematics.
These include the line-of-sight velocities of neutral and ionized
hydrogen clouds with their distances estimated by the tangential
point method (Clemens 1985; McClure-Griffiths and Dickey 2007;
Levine et al. 2008), Cepheids with the distance scale based on the
period–luminosity relation, open star clusters and OB associations
with photometric distances (Mishurov and Zenina 1999; Rastorguev
et al. 1999; Zabolotskikh et al. 2002; Bobylev et al. 2008;
Mel’nik and Dambis 2009), and masers with their trigonometric
parallaxes measured by VLBI (Reid et al. 2009; McMillan and Binney
2010; Bobylev and Bajkova 2010; Bajkova and Bobylev 2012). The
youngest massive high luminosity (OB) stars are also of great
interest, because they have not moved far away from their
birthplace in their lifetime, clearly tracing the Galactic spiral
structure. Such stars were used by many authors on various spatial
scales to study the Galactic kinematics (Moffat et al. 1998;
Avedisova 2005; Popova and Loktin 2005; Branham 2006). Among the
youngest OB stars, the fraction of binary and multiple systems
reaches 70--80\%. However, the kinematics of precisely such
systems has been studied more poorly than that of single stars.

At present, great interest in massive binaries is related to the
study of the Galaxy’s spatial, kinematic, and dynamical
properties. For example, it is important to know the positions of
high-mass X-ray binaries in the Galaxy (Coleiro and Chaty 2012),
their connection with the Galactic spiral structure (Lutovinov et
al. 2005), the distribution of possible presupernovae, the
progenitors of neutron stars and black holes, around the Sun
(Hohle et al. 2010), the distribution of runaway OB stars
(Tetzlaff et al. 2011), etc.

Binary systems usually have a long history of spectroscopic and
photometric observations. Therefore, in particular, their systemic
line-of-sight velocities, spectral classification, and photometry
are well known. Here, we focus on revealing the binary systems
containing predominantly O stars that are suitable for studying
the kinematic characteristics of the solar neighborhood and the
Galaxy as a whole. Among the O stars from the Hipparcos catalogue
(1997), only 12 have trigonometric parallaxes differing
significantly from zero and the farthest of them, 10 Lac, is only
$\approx$540 pc away from the Sun (Maiz-Apell\'aniz et al. 2008).
Therefore, we use less reliable spectrophotometric distance
estimates for many of the stars from our sample.

The goal of this paper is to create a kinematic database on
Galactic massive $(>10M_\odot)$ young star systems within 2--3~kpc
of the Sun, to determine the main kinematic parameters of the
Galaxy, including the spiral density wave parameters, using this
database, to reveal runaway stars, to explain the kinematic
peculiarities of the (Gould Belt) stars nearest to the Sun, and to
study the influence of the latter on the Galactic parameters being
determined.

\section{DATA}
\subsection{Distances}
Studying the total space velocities of stars requires knowing six
quantities: the equatorial coordinates ($\alpha,\delta$), the
heliocentric distance ($r$), two proper motion components
($\mu_\alpha\cos\delta, \mu_\delta$), and the line-of-sight
velocity ($V_r$). It is important to note that the requirement of
a high accuracy for the space velocities imposes a constraint on
the heliocentric distances of stars, which is no more than 3~kpc
in our case. For example, since $V_t=4.74 r \mu_t,$ where
$\mu_t=\sqrt{\mu^2_\alpha\cos^2\delta+\mu^2_\delta },$ at a
typical error $\sigma_{\mu_t}\approx1.4$~mas yr$^{-1},$ the
contribution from the proper motion error to the total velocity of
a star will be a significant fraction of the total velocity,
$\sigma_{V_t}\approx20$~km s$^{-1},$ already at a distance of
3~kpc.

Our sample was produced as follows.

First of all, it includes O-type stars for which the relative
error of the trigonometric parallax does not exceed $\approx$15\%.
Among these stars, there are also single ones. The distances
corrected for the Lutz--Kelker bias were taken from
Maiz-Apell\'aniz et al. (2008).

Our sample includes five stars whose components are massive O
stars with known highly accurate orbital (dynamical) parallaxes:
 $\theta^1$Ori C (Kraus et al. 2009),
 $\sigma$~Ori (Caballero 2008),
 $\gamma^2$Vel (North et al. 2007a),
 $\delta$~Sco (Tycner et al. 2011), and
 $\sigma$~Sco (North et al. 2007b). For four of these systems (except $\sigma$~Ori), the
orbital semimajor axes were measured with ground-based optical
interferometers, such as, for example, SUSI (Sydney University
Stellar Interferometer).

An example of a high accuracy of the dynamical parallax is the
triple system
 $\theta^1$Ori C (Sp: O+?+?; with a total mass of $44M_\odot$).
It is one of the Orion Trapezium members for which the Hipparcos
trigonometric parallax is negative. In contrast, the distance
calculated from the dynamical parallax is $r=410\pm20$~pc. This is
in excellent agreement with the estimate of $418\pm6$~pc to the
region Orion--KL obtained by VLBI within the VERA program (Japan)
using the maser SiO emission at 43.122 GHz (Kim et al. 2008). For
the remaining six systems from this list
 $\sigma$~Ori (O9.5+B0.5; $20M_\odot+12M_\odot$),
 $\gamma^2$Vel (O+WR; $29M_\odot+9M_\odot$),
 $\delta$~Sco (B1+B1; $12M_\odot+7M_\odot$),
 $\sigma$~Sco (B0+B1; $18M_\odot+12M_\odot$),
 $\beta$~Cen  (B1+B1; $11M_\odot+10M_\odot$),
 $\lambda$~Sco (B1.5+B2+PMS; $10M_\odot+8M_\odot$+?), the situation is less
dramatic—their dynamical parallaxes have smaller errors (the
trigonometric distances were improved by analyzing the orbits).
The distances calculated using the dynamical parallaxes for
$\beta$~Cen and $\lambda$~Sco make them twice as close compared to
the trigonometric estimates (Ausseloos et al. 2006; Tango et al.
2006).

Among the visual binary stars in our sample, there are two massive
triple systems, $\beta$~Cep (WDS J21287+7034) and $\gamma$~Lup
(WDS J15351--4110), with known estimates of their orbital
parallaxes (Docobo and Andrade 2006) that are as accurate as the
trigonometric measurements. Significantly, for example, for
$\gamma$~Lup (B1+B2+B3; 12M$_\odot$+8M$_\odot$+7M$_\odot$), the
dynamical parallax $\pi_f=6.6\pm0.4$~mas was estimated at the then
available trigonometric parallax estimate
$\pi_{tr}=5.75\pm1.24$~mas (Hipparcos 1997). In contrast, the
trigonometric parallax from the revised version of Hipparcos is
$\pi_{tr}=7.75\pm0.50$~mas (van Leeuwen 2007); it became larger in
agreement with the analysis of the stellar orbits. Similarly, for
$\beta$~Cep (B3+A7+A0; 19M$_\odot$+2M$_\odot$+3M$_\odot$), it was
required to reduce the trigonometric parallax, which was
subsequently confirmed.

The unique system V729~Cyg, which also enters into our sample, is
not only an optical star but also a radio one. According to
Kennedy et al. (2010) and Dzib et al. (2012), it consists of four
components with a total mass from 50M$_\odot$ to 90M$_\odot$
(component A is a spectroscopic binary). Its trigonometric
parallax (with an error $\sigma_\pi/\pi\approx$30\%) and proper
motion components were measured by VLBI and its dynamical parallax
was estimated (Dzib et al. 2012). Based on these estimates, we
adopted the distance $r=1500\pm300$~pc for it. This system is
interesting to us, because the previously available Hipparcos
proper motions and the line-of-sight velocity $V_r=-33\pm4$~km
s$^{-1}$ (Rauw et al. 1999) showed it to be a runaway star (see
Section 3.5) with a peculiar velocity $|V_{pec}|\approx50$~km
s$^{-1}$. At the new parameters from Kennedy et al. (2010)
($V_r=-5.9\pm4.7$~km s$^{-1}$) and Dzib et al. (2012) that we
adopted, its peculiar velocity is $|V_{pec}|\approx25$~km
s$^{-1}$, i.e., it is a runaway star.

%%%%%%%%%%%%%%
 \begin{table}[p]                                     % T~1.
 \caption[]{\small Information about the published sources }
 \begin{center}
 \label{t:01}
 \small
 \begin{tabular}{|r|c|r|c|}\hline

     Star & $V_\gamma$, $~\mu,~$ $dist$&    Star & $V_\gamma$, $~\mu,~$ $dist$ \\\hline

 $\delta$ Ori A   &  Sb9, UC4, HpA  &   XZ Cep       & Sb9, L07, H97   \\
 $\delta$ Ori C   &  Sb9, UC4, HpA  &   DH Cep       & Sb9, L07, Hi96  \\
 $\lambda$ Ori A  &  G06, UC4, HpA  &   LZ Cep       & Sb9, L07, G12   \\
 $\theta^2$ Ori A &  Sb9, UC4, HpA  &   HD  48099    & Sb9, L07, Mh10  \\
 $\zeta$ Ori A    & CRV2, UC4, HpA  &   HD 152218    & Sb9, UC4, S08   \\
 15 Mon           & Cv10, UC4, HpA  &   HD 152248    & Sb9, UC4, S01   \\
 10 Lac           & CRV2, UC4, HpA & ~~CPD$-$41 7733 & Sb9, UC4, S07   \\
 $\theta^1$ Ori C &  Sb9, UC4, Dyn1 &   HD  15558    & Sb9, L07, G12   \\
 $\sigma$ Ori     &  Sb9, UC4, Dyn2 &   HD  35921    & Sb9, L07, L07   \\
 $\gamma^2$ Vel   &  G06, L07, Dyn3 &   HD  53975    & CRV2, L07, G12  \\
 $\delta$ Sco     &  Sb9, L07, Dyn4 & $\tau$ CMa     & Sb9, L07, G12   \\
 $\sigma$ Sco     &  Sb9, L07, Dyn5 &   HD  75759    & Sb9, L07, L07   \\
 $\beta$ Cen      & Au06, L07, Dyn6 &    V1104 Sco   & Ri11, Tyc2, Mr07 \\
 $\lambda$ Sco    & Uy04, L07, Dyn7 &    HD 37737    & Mc07,  L07, Mc07 \\
 $\beta$ Cep      &  Sb9, L07, L07  &   HD 229234    & Mh13, UC4, COCD \\
 $\gamma$ Lup     &  Sb9, L07, L07  &   HD   1337    &  Sb9, L07, G12  \\
 AG Per           &  Sb9, L07, T10  &   QZ Car       & Sb9, L07, G12   \\
 V1174 Ori        & Sb9, PMXL, T10  & $\delta$ Cir   & Sb9, L07, G12   \\
 V1388 Ori        &  Sb9, L07, T10  &   HD 167771    & Sb9, L07, L85   \\
 V578 Mon         &  Sb9, UC4, T10  &   HD 191201    & Sb9, L07, G12   \\
 RS Cha           &  Sb9, L07, T10  &   HD 199579    &  Sb9, L07, G12  \\
 CV Vel           &  Sb9, L07, T10  & Plaskett star  &  Sb9, L07, G12  \\
 QX Car           &  Sb9, L07, T10  &   HD 36822     &  Sb9, L07, L07  \\
 EM Car           &  Sb9, UC4, T10  &   HD  96670    &  Sb9, L07, G12  \\
 V760 Sco         &  Sb9, L07, T10  &   HD 101205    & CRV2, L07, G12  \\
 U Oph            &  Sb9, L07, T10  & $\theta$ Mus   &  Sb9, L07, G12  \\
 V539 Ara         &  Sb9, L07, T10  &   HD 154368    & CRV2, L07, G12  \\
 V3903 Sgr        &  Sb9, UC4, T10  &   HD 159176    &  Sb9, L07, G12  \\
 DI Her           &  Sb9, L07, T10  &   HD 165052    &  Sb9, L07, G12  \\
 V453 Cyg         &  Sb9, CRV2, T10 &   HD 206267    &  Sb9, L07, G12  \\
 V478 Cyg         &  Sb9, L07, T10  &   HD 152147    & Wi13, UC4, Wi13 \\
 AH Cep           &  Sb9, L07, T10  & BD$-$16 4826   & Wi13, UC4, Wi13 \\
 CW Cep           &  Sb9, L07, T10  &   HD 193443    & Mh13, L07, G12  \\
 DW Car           &  Sb9, UC4, T10  &   9 Sgr        &  R12, L07, CaII \\
 V1034 Sco        &  Sb9, UC4, T10  & $\gamma$ Cas   & SIMB, UC4, CaII \\
  V961 Cen        &  Sb9, L07, P02  &   HD  93403    &  Sb9, L07, CaII \\
 ~1H 1249$-$637   & CRV2, L07, L07  &   HD  57060    &  Sb9, L07, CaII \\
 Cyg X-1          &  Sb9, L07, Z05  &   V641 Mon     & CRV2, L07, G12  \\

 \hline
 \end{tabular}
 \end{center}
 \end{table}
%%%%%%%%%%%%%%
  \begin{table*}[]
  \centerline {Table 1. Contd.}
  \begin{center}
  \small
  \begin{tabular}{|r|c|r|c|}\hline

      Star & $V_\gamma$, $~\mu,~$ $dist$&    Star & $V_\gamma$, $~\mu,~$ $dist$ \\\hline

 SZ Cam           &  Sb9, UC4, L98  &   FZ CMa       &  Sb9, L07, G12  \\
 TT Aur           &  Sb9, UC4, W86  &   V599 Aql     &  Sb9, L07, L07  \\
 IU Aur           &  Sb9, L07, L07  &   NY Cep       &  Sb9, L07, H90  \\
 LT CMa           &  Sb9, L07, B10  &   CV Ser       &  Sb9, L07, G12  \\
 V Pup            &  Sb9, L07, L07  &   HD  46149    & Mh09, L07, Mh09 \\
 AI Cru           &  Sb9, L07, B87  &   $o$ Ori      &  Sb9, L07, L07  \\
 VV Ori           &  Sb9, L07, TM07 &   HD  93130    &  Sb9, UC4, G12  \\
 V716 Cen    &  Sb9, L07, B08  &         WR  79    &   Sb9, L07, Hu01 \\
 ~~V448 Cyg  &  Sb9, L07, H97  &         WR 139    &   Sb9, L07, Hu01 \\
 $\mu^1$ Sco &  Sb9, L07, L07  & ~~~~~~~~V1292 Sco &   Sb9, UC4, S06  \\
 u Her       &  Sb9, L07, L07  &         TU Mus    &   Sb9, UC4, G12  \\
 HD  1383    &  Sb9, L07, G12  &         HR 1952   &   Sb9, L07, L07  \\
 HR 6187     & Mh12, UC4, Mh12 &         V729 Cyg  & Kn10, Dz12, Dz12 \\
 \hline
 \end{tabular}
 \end{center}
 {\small
Notes.

1. The sources of line-of-sight velocities: Sb9 (Pourbaix et al.
2004); Cv10 (Cvetkovi\'c et al. 2010); CRV2 (Kharchenko et al.
2007); G06 (Gontcharov 2006); Ri11 (Richardson et al. 2011); Mc07
(McSwain et al. 2007); Wi13 (Williams et al. 2013); Uy04
(Uytterhoeven et al. 2004); Mh09, Mh10, Mh12, Mh13 (Mahy et al.
2009, 2010, 2012, 2013); R12 (Rauw et al. 2012); Kn10 (Kennedy et
al. 2010); SIMB (SIMBAD database).

2. The sources of proper motions: L07 (van Leeuwen 2007); UC4
(Zacharias 2012); Tyc2 (Hog et al. 2000); PMXL (Roeser et al.
2010).

3. The sources of distances: HpA (Maiz-Apell\'aniz et al. 2008);
T10 (Torres et al. 2009); Dyn1 (Kraus et al. 2009); Dyn2
(Caballero 2008); Dyn3 (North et al. 2007a); Dyn4 (Tycner et al.
2011); Dyn5 (North et al. 2007b); Dyn6, Au06 (Ausseloos et al.
2006); Dyn7 (Tango et al. 2006); G12 (Gudennavar et al. 2012);
CaII (Megier et al. 2009); Z05 (Zi\'o\l kowski 2005); L85
(Lindroos 1985); L98 (Lorenz et al. 1998); W86 (Wachmann et al.
1986); B87 (Bell et al. 1987); B08, B10 (Bakis et al. 2008, 2010);
Hu01 (van der Hucht 2001); H90 (Holmgren et al. 1990); Hi96
(Hilditch et al. 1996); H97 (Harries et al. 1997); S01, S06, S07,
S08 (Sana et al. 2001, 2006, 2007, 2008); COCD~(Kharchenko et al.
2005); P02 (Penny et al. 2002); Mr07 (Marcolino et al. 2007); TM07
(Terrell et al. 2007); Dz12 (Dzib et al. 2012).
 }
      \end{table*}
%%%%%%%%%%%%%%%%%%%%%%%%%%%%%%%%%%%%%%%%%%%%%%%%%%%%%%%%%%%%%%%%%
%%%%%%%%%%%%%%
 \begin{table}[t]                                     % T~2.
 \caption[]{\small Numbers of the Hipparcos B stars used}
 \begin{center}
 \label{t:02}
 \small
 \begin{tabular}{|r|r|r|r|r|r|r|r|r|r|r|r|}\hline
   1067 &   2903 &   2920 &   4427 &   8068 &   8886 &  12387 &  17313 &  18081 &  18246 \\
  18434 &  18532 &  21060 &  21444 &  22549 &  23734 &  23972 &  24845 &  25223 &  25336 \\
  25539 &  26248 &  27364 &  27366 &  27658 &  28199 &  28237 &  29276 &  29417 &  29771 \\
  29941 &  30324 &  30772 &  31125 &  32463 &  32759 &  33092 &  33575 &  33579 &  33971 \\
  35037 &  35083 &  35363 &  35855 &  36362 &  38010 &  38020 &  38438 &  38500 &  38518 \\
  38593 &  38872 &  39961 &  39970 &  40274 &  40285 &  40321 &  40357 &  41296 &  42568 \\
  42828 &  43937 &  45080 &  45122 &  45941 &  50067 &  52633 &  54327 &  59196 &  59747 \\
  60009 &  60718 &  61585 &  62434 &  63003 &  64004 &  66657 &  67464 &  67472 &  68245 \\
  68282 &  68862 &  69996 &  70300 &  70574 &  71121 &  71352 &  71860 &  71865 &  73273 \\
  73334 &  77635 &  77840 &  77859 &  78384 &  78820 &  78918 &  78933 &  80473 &  80569 \\
  80911 &  81266 &  82110 &  82545 &  85267 &  85696 &  86670 &  88149 &  88714 &  88886 \\
  91918 &  92133 &  92609 &  92728 &  92855 &  93299 &  99303 &  94481 &  97679 & 100751 \\
 101138 & 103346 & 106227 & 114104 &        &        &        &        &        &        \\
 \hline
 \end{tabular}
 \end{center}
 \end{table}
%%%%%%%%%%%%%%

The detached spectroscopic binaries with known spectroscopic
orbits selected by Torres et al. (2010) on condition that the
masses and radii of both components are determined with errors of
no more than $\pm3\%$ constitute a no less important part of our
sample. These stars served to check fundamental relations, such as
the mass--radius and mass--luminosity ones, to test stellar
evolution models, etc. It is important that their spectroscopic
distances agree with the trigonometric ones (van Leeuwen 2007)
within $<10\%$ error limits (which was calculated from an
overlapping sample). We took all youngest ($<50$~Myr) and most
massive (the sum of the components $>10M_\odot$) systems from the
list by Torres et al. (2010). There is one exception---RS Cha is a
young ($\approx8$~Myr) system, but both its components are
low-mass ones ($\approx1M_\odot$, these are T~Tau stars).

We also used the list of massive spectroscopic binaries by Hohle
et al. (2010) and the list of spectroscopic binaries by Harries et
al. (1997). Our sample includes three stars from the list of
high-mass X-ray binaries with distance estimates by Coleiro and
Chaty (2012): $\gamma$ Cas, 1H 1249$-$637, and Cyg X-1. The
distances for a number of stars were taken from the compilation by
Gudennavar et al. (2012); we set the distance error equal to 20\%.
It can be seen from a number of stars from this compilation that
the distance estimates tend to decrease, sometimes significantly,
depending on the publication time.

Note that previously (Bobylev and Bajkova 2011) we studied the
Galactic kinematics using OB3 stars ($r>0.8$~kpc) whose distances
were estimated (Megier et al. 2009) from interstellar CaII
absorption lines. In this paper, we essentially managed to avoid
the overlap between the samples, because the results obtained in
different distance scales are of interest. As can be seen from
Table 1, where the entire bibliography is reflected, only four
stars have distances in the calcium scale.

Finally, we selected all Hipparcos stars of spectral types from B0
to B2.5 (according to the stellar evolution models, the masses of
such stars exceed $10-7M_\odot$) whose parallaxes were determined
with errors of no more than 10\% and for which there are
line-of-sight velocities in the catalog by Gontcharov (2006). Note
that at present, such a sample is easy to draw from the catalog by
Anderson and Francis (2012). There are 124 such B stars; their
Hipparcos numbers are given in Table 2.

As a result, we obtained a sample of 220 stars whose distribution
in the Galactic plane is shown in Fig.~1a. More than two thirds of
the entire sample or, more precisely, 162 stars are located within
about 600~pc of the Sun. Almost all of them belong to the Gould
Belt (Bobylev 2006; Bobylev and Bajkova 2007). This membership is
suggested by the characteristic inclination of the stars from the
selected neighborhood to the Galactic plane ($\approx17^\circ$),
which can be seen from Fig.~1b. The distant stars (Fig. 1a)
clearly delineate the Perseus (farther from the Sun) and
Carina--Sagittarius (closer to the Galactic center) spiral arms;
the Orion arm is well represented (the elongation in a direction
$l\approx75^\circ$). In our view, the distribution of stars shown
in Fig.~1a is as detailed as the distribution map of OB stars from
Patriarchi et al. (2003), where the distances were determined from
the 2MASS photometric data.

  \begin{figure}[t]
 {\begin{center}
  \includegraphics[width=150mm]{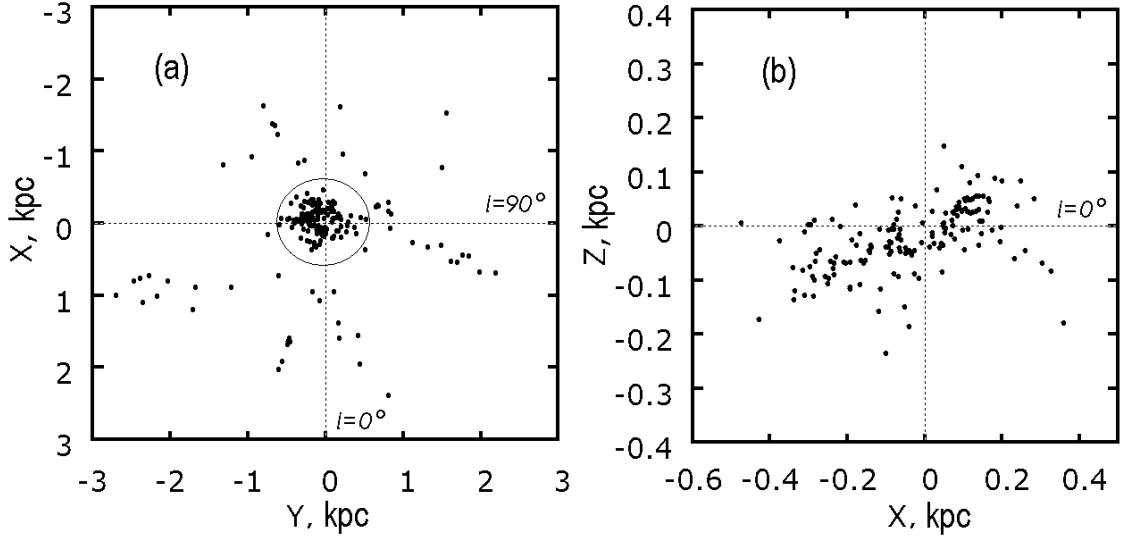}
  \caption{\small
(a) Distribution of stars on the Galactic XY plane; the circle
marks the neighborhood the distribution of stars from which is
shown on the XZ plane on panel~(b).
 }
 \label{f-XY}
 \end{center}}
 \end{figure}

\subsection{Line-of-Sight Velocities}
In addition to the main source, the Sb9 bibliographic database on
the orbits of spectroscopic binaries (Pourbaix et al. 2004), we
used the most recent orbit determinations for a number of stars,
for example, the results from Rauw et al. (2012), Williams et al.
(2013), and Mahy et al. (2013).

Note that the SIMBAD search database sometimes contains
significant inaccuracies with regard to the line-of-sight
velocities for spectroscopic binaries. For example,
$V_r=-101\pm20$~km s$^{-1}$ for the star V1034~Sco (O9.5+B1.5)
from the list by Torres et al. (2010), while, according to Sana et
al. (2005), $V_\gamma=-16.0\pm1.5$~km s$^{-1}$.

After checking against the Sb9 database, we took the line-of-sight
velocities of all spectroscopic binaries from the latest
publications. Therefore, there are significant deviations from the
data of the Gontcharov (2006) or CRVAD-2 (Kharchenko et al. 2007)
catalogs for many of them.

As was pointed out in Moffat et al. (1998), determining realistic
values of the systemic line-of-sight velocities is very difficult
for O stars and virtually impossible for Wolf–Rayet stars.
Therefore, we tried to include as few data on Wolf–Rayet stars as
possible in our sample.

\subsection{Proper Motions}
We took most of the stellar proper motions from the Hipparcos
catalogue (van Leeuwen 2007). For 17 systems that do not enter
into the Hipparcos list, we used data from the UCAC4 (Zacharias
2012), Tycho-2 (Hog et al. 2000), CRVAD-2 (Kharchenko et al.
2007), and PPMXL (Roeser et al. 2010) catalogues.

Thus, O stars, for example, almost all spectroscopic binary O
stars from the GOS2 catalogue (Maiz-Apell\'aniz et al. 2004) are
reflected in our sample to a maximum degree. Only the stars with
the ``runaway'' status were rejected.

\section{KINEMATIC ANALYSIS}
\subsection{Galactic Rotation Model}
The method of determining the kinematic parameters applied here
consists in minimizing a quadratic functional F:
  \begin{equation}
  \min~F=\sum_{j=1}^N w_r^j (V_r^j-\hat{V}_{r}^j)^2
        +\sum_{j=1}^N w_l^j (V_l^j-\hat{V}_{l}^j)^2
        +\sum_{j=1}^N w_b^j (V_b^j-\hat{V}_{b}^j)^2
  \label{Functional}
\end{equation}
provided the fulfilment of the following constraints derived from
Bottlinger’s formulas with an expansion of the angular velocity of
Galactic rotation $\Omega$ into a series to terms of the second
order of smallness with respect to $r/R_0$ and with allowance made
for the influence of the spiral density wave:
\begin{equation}
 \begin{array}{lll}
 V_r&=&-u_\odot\cos b\cos l-v_\odot\cos b\sin l-w_\odot\sin b\\
 &+&R_0(R-R_0)\sin l\cos b \Omega'_0\\
 &+&0.5R_0 (R-R_0)^2 \sin l\cos b \Omega''_0\\
 &+&\Delta V_{rot}\sin(l+\theta)\cos b,\\
 &-&V_R \cos(l+\theta)\cos b,
 \label{EQ-1}
 \end{array}
 \end{equation}
 \begin{equation}
 \begin{array}{lll}
 V_l&=& u_\odot\sin l-v_\odot\cos l\\
  &+&(R-R_0)(R_0\cos l-r\cos b) \Omega'_0\\
  &+&(R-R_0)^2 (R_0\cos l - r\cos b)0.5\Omega''_0 - r \Omega_0 \cos b\\
  &+&\Delta V_{rot} \cos(l+\theta)+V_R\sin(l+\theta),
 \label{EQ-2}
 \end{array}
 \end{equation}
 \begin{equation}
 \begin{array}{lll}
 V_b&=&u_\odot\cos l \sin b + v_\odot\sin l \sin b-w_\odot\cos b\\
    &-&R_0(R-R_0)\sin l\sin b\Omega'_0\\
    &-&0.5R_0(R-R_0)^2\sin l\sin b\Omega''_0\\
    &-&\Delta V_{rot} \sin (l+\theta)\sin b,\\
    &+&V_R \cos(l+\theta)\sin b,
 \label{EQ-3}
 \end{array}
 \end{equation}
where the following notation is used: $N$ is the number of stars
used; $j$ is the current star number; $V_r$ is the line-of-sight
velocity, $V_l=4.74 r \mu_l\cos b$ and $V_b=4.74 r \mu_b$ are the
proper motion velocity components in the $l$ and $b$ directions,
respectively, with the coefficient 4.74 being the quotient of the
number of kilometers in an astronomical unit and the number of
seconds in a tropical year;
 $\hat{V}_{r}^j, \hat{V}_{l}^j, \hat{V}_{b}^j$
are the measured components of the velocity field (data);
 $w_r^j, w_l^j, w_b^j$
are the weight factors; the star's proper motion components
$\mu_l\cos b$ and $\mu_b$ are in mas yr$^{-1}$ and the
line-of-sight velocity $V_r$ is in km s$^{-1}$;
$u_\odot,v_\odot,w_\odot$ are the stellar group velocity
components relative to the Sun taken with the opposite sign (the
velocity $u$ is directed toward the Galactic center, $v$ is in the
direction of Galactic rotation, $w$ is directed to the north
Galactic pole); $R_0$ is the Galactocentric distance of the Sun;
$R$ is the Galactocentric distance of the star; $\Omega_0$ is the
angular velocity of rotation at the distance $R_0$; the parameters
$\Omega'_0$ and $\Omega''_0$ are the first and second derivatives
of the angular velocity, respectively; the distance $R$ is
calculated from the formula
 \begin{equation}
 \begin{array}{lll}
 R^2=r^2\cos^2 b-2R_0 r\cos b\cos l+R^2_0.
 \label{RR}
 \end{array}
 \end{equation}
To take into account the influence of the spiral density wave, we
used the simplest kinematic model based on the linear density wave
theory by Lin et al. (1969), in which the potential perturbation
is in the form of a traveling wave. Then,
 \begin{equation}
 \begin{array}{lll}
      V_R=&f_R \cos \chi,
 \label{VR-VR}
 \end{array}
 \end{equation}
 \begin{equation}
 \begin{array}{rll}
      \Delta V_{rot}=&f_\theta \sin \chi,
 \label{VR-Vtheta}
 \end{array}
 \end{equation}
where $f_R$ and $f_\theta$ are the amplitudes of the radial
(directed toward the Galactic center in the arm) and azimuthal
(directed along the Galactic rotation) velocity perturbations; $i$
is the spiral pitch angle ($i<0$ for winding spirals); $m$ is the
number of arms, we take $m=2$ in this paper; $\theta$ is the
star’s position angle (measured in the direction of Galactic
rotation): $\tan\theta = y/(R_0-x)$, where $x$ and $y$ are the
Galactic heliocentric rectangular coordinates of the object; the
wave phase $\chi$ is
 \begin{equation}
   \chi=m[\cot (i)\ln (R/R_0)-\theta]+\chi_\odot,
 \label{chi-creze}
 \end{equation}
where $\chi_\odot$ is the Sun's radial phase in the spiral density
wave; we measure this angle from the center of the
Carina--Sagittarius spiral arm ($R\approx7$~kpc). The parameter
$\lambda$, the distance (along the Galactocentric radial
direction) between adjacent segments of the spiral arms in the
solar neighborhood (the wavelength of the spiral density wave), is
calculated from the relation
\begin{equation}
 \frac{2\pi R_0}{\lambda} = m\cot(i).
 \label{a-04}
\end{equation}
The weight factors in functional (1) are assigned according to the
following expressions (for simplification, we omitted the index
$i$)
 \begin{equation}
 w_r=S_0/\sqrt {S_0^2+\sigma^2_{V_r}},\quad
 w_l=\beta^2 S_0/\sqrt {S_0^2+\sigma^2_{V_l}},\quad
 w_b=\gamma^2 S_0/\sqrt {S_0^2+\sigma^2_{V_b}},
 \label{WESA}
 \end{equation}
where $S_0$ denotes the dispersion averaged over all observations,
which has the meaning of a ``cosmic'' dispersion taken to be 8~km
s$^{-1}$; $\beta=\sigma_{V_r}/\sigma_{V_l}$ and
$\gamma=\sigma_{V_r}/\sigma_{V_b}$ are the scale factors that we
determined using data on open star clusters (Bobylev et al. 2007),
$\beta=1$ and $\gamma=2$. The errors of the velocities $V_l$ and
$V_b$ are calculated from the formula
 \begin{equation}
 \sigma_{(V_l,V_b)}=4.74r\sqrt{\mu^2_{l,b}\Biggl({\sigma_r\over r}\Biggr)^2+\sigma^2_{\mu_{l,b}}}.
 \label{Errors}
 \end{equation}
The optimization problem (1)–(8) is solved for the unknown
parameters $u_\odot,$ $v_\odot,$ $w_\odot,$ $\Omega_0$,
$\Omega'_0,$ $\Omega''_0,$ $f_R,$ $f_\theta,$ $i$ and $\chi_\odot$
by the coordinate-wise descent method.

We estimated the errors of the sought-for parameters through Monte
Carlo simulations. The errors were estimated by performing 1000
cycles of computations. For this number of cycles, the mean values
of the solutions essentially coincide with the solutions obtained
from the input data without any addition of measurement errors.
Measurements errors were added to such input data as the
line-of-sight velocities, proper motions, and distances.

We take $R_0=8.0\pm0.4$~kpc according to the result of analyzing
the most recent determinations of this quantity in the review by
Foster and Cooper (2010).

\subsection{Periodogram Analysis of Residual Velocities}
Relation (8) for the phase can be expressed in terms of the
perturbation wavelength $\lambda$, which is equal to the distance
between the adjacent spiral arms along the Galactic radius vector.
Using relation (9) between the pitch angle $i$ and wavelength
$\lambda$ (9), we will obtain an expression for the phase as a
function of the star’s Galactocentric distance $R$ and position
angle $\theta$:
\begin{equation}
 \chi = \frac {2\pi R_0}{\lambda} \ln(R/R_0)- m\theta+\chi_{\odot}.
 \label{a-05}
 \end{equation}
The goal of our spectral analysis of the series of measured
residual velocities $V_{R_n}$ and $\Delta V_{\theta_n},
n=1,\dots,N$, where $N$ is the number of objects, is to extract
the periodicity in accordance with model (6), (7), describing a
spiral density wave with parameters $f_R, f_\theta, \lambda$ and
$\chi_\odot$. If the wavelength $\lambda$ is known, then the pitch
angle $i$ is easy to determine from Eq. (9) by specifying the
number of arms $m.$ Here, we adopt a two-armed model, i.e., $m=2.$

Let us form the initial series of velocity perturbations (6), (7)
for Galactic objects in the most general, complex form:
\begin{equation}
 V_n=V_{R_n}+j\Delta V_{\theta_n},
 \label{a-06}
\end{equation}
where $j=\sqrt-1$, $n$ is the object number ($n=1, . . . ,N).$ A
periodogram analysis consists in calculating the amplitude of the
spectrum squared (power spectrum) obtained by expanding series
(13) in terms of orthogonal harmonic functions
 $\exp[-j (2\pi R_0/\lambda_k)\ln(R_n/R_0)+j m\theta_n]$
in accordance with Eq. (12) for the phase of the spiral density
wave to extract significant peaks in it.

Let us first calculate the complex spectrum of our series:
 \begin{equation}
 \renewcommand{\arraystretch}{1.8}
 \begin{array}{lll}
 \bar{V}_{\lambda_k}=  \bar{V}_{\lambda_k}^{Re}+j\bar{V}_{\lambda_k}^{Im}= \\
 \frac{1}{N} \sum_{n=1}^N (V_{R_n}+j\Delta V_{\theta_n}) \exp[-j\frac{\displaystyle 2\pi R_0}{\displaystyle \lambda_k}\ln(R_n/R_0)+j m\theta_n],
 \label{a-07}
 \end{array}
 \end{equation}
where the superscripts $Re$ and $Im$ denote the real and imaginary
parts of the spectrum.

Let us reduce the problem of calculating spectrum (14) to the
standard Fourier transform. Obviously, transformation (14) can be
represented as
 \begin{equation}
 \renewcommand{\arraystretch}{1.8}
 \begin{array}{lll}
 \bar{V}_{\lambda_k}=\frac{1}{N}\sum_{n=1}^N (V_{R_n}+ j\Delta V_{\theta_n})\exp(j{\rm m}\theta_n)
      \exp[-j\frac{\displaystyle 2\pi R_0}{\displaystyle \lambda_k}\ln(R_n/R_0)]= \\
 =\frac{1}{N}\sum_{n=1}^N V_n^{'}  \exp[-j\frac{\displaystyle 2\pi R_0}{\displaystyle \lambda_k}\ln(R_n/R_0)],
 \label{a-08}
 \end{array}
 \end{equation}
where
 \begin{equation}
 \renewcommand{\arraystretch}{1.6}
 \begin{array}{lll}
 V_n^{'}=V_n^{Re}+jV_n^{Im}=\\
 =[V_{R_n}\cos(m \theta_n)-\Delta V_{\theta_n}\sin(m \theta_n)]+
 j[V_{R_n}\sin(m \theta_n)+\Delta V_{\theta_n}\cos( m \theta_n)].
 \label{a-09}
 \end{array}
 \end{equation}
Finally, making the change of variables
\begin{equation}
 R_n^{'}=R_0\ln(R_n/R_0),
 \label{a-10}
\end{equation}
we will obtain the standard Fourier transform of the sequence
$V_n^{'}$, precalculated from Eq. (16) and defined in the space of
coordinates $R_n^{'}$ (17):
\begin{equation}
 \bar{V}_{\lambda_k}=\frac{1}{N}\sum_{n=1}^N
 V_n^{'}\exp[-j\frac{2\pi R_n^{'}}{\lambda_k}].
 \label{a-11}
\end{equation}
The periodogram $|\bar{V}_{\lambda_k}|^2$ from which the
statistically significant peak whose coordinate determines the
sought-for wavelength $\lambda$ and, accordingly, the pitch angle
of the spiral density wave $i$ (see (9)) is extracted is to be
analyzed further. Obviously, our spectral analysis of the complex
sequence (13) composed of the radial and tangential residual
velocities does not allow the radial and tangential perturbation
amplitudes to be separated. A separate analysis of the radial,
$\{V_{R_n}\}$, and tangential residual, $\{\Delta V_{\theta_n}\}$,
velocities, for example, as was shown in Bajkova and Bobylev
(2012), is required to estimate the perturbation amplitudes $f_R$
and $f_\theta$. The input data  $V_{n}^{'}(R^{'}) (n=1,\dots,N)$
should be specified on a discrete mesh $l=1,\dots,K=2^\alpha$,
$\alpha$ is an integer $>0$, $N\le K$; $\Delta_R$, for a numerical
realization of the Fourier transform (18), for example, using fast
algorithms. The coordinates of the data are defined as numbers
rounded off to an integer as follows:
$l_n\approx[(R_n^{'}+|\min\{R_k^{'}\}|_{k=1,...,N})/\Delta_R+1],
n=1,...,N$. is the discretization interval. The specified sequence
of data is assumed to be periodic, with its period being
$D=K\times\Delta_R$. The perturbation amplitudes are assumed to be
zero at the points $l$ where there are no measurements.

  \begin{figure}[t]
 {\begin{center}
  \includegraphics[width=150mm]{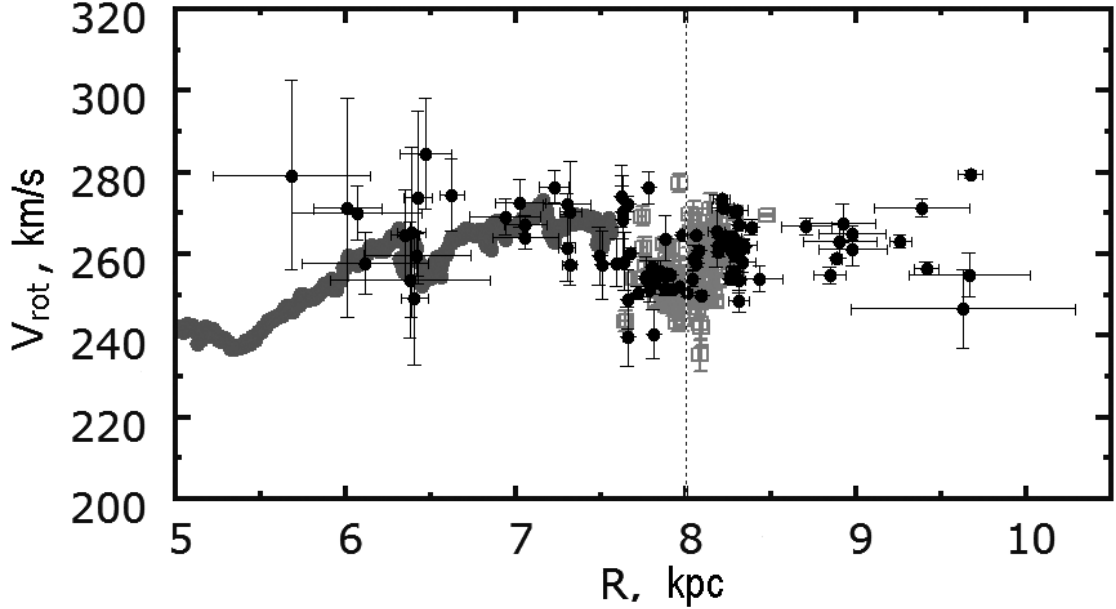}
  \caption{\small
Rotation velocities of stars around the Galactic center $V_{rot}$
versus Galactocentric distance $R$: the thick line indicates the
velocities of neutral hydrogen (McClure-Griffiths and
Dickey~2007); the black circles with bars are massive
spectroscopic binaries; the grays squares are massive Hipparcos B
stars whose parallaxes were determined with errors of no more than
10\%.
 }
 %%\label{f-XY}
 \end{center}}
 \end{figure}

  \begin{figure}[p]
 {\begin{center}
  \includegraphics[width=150mm]{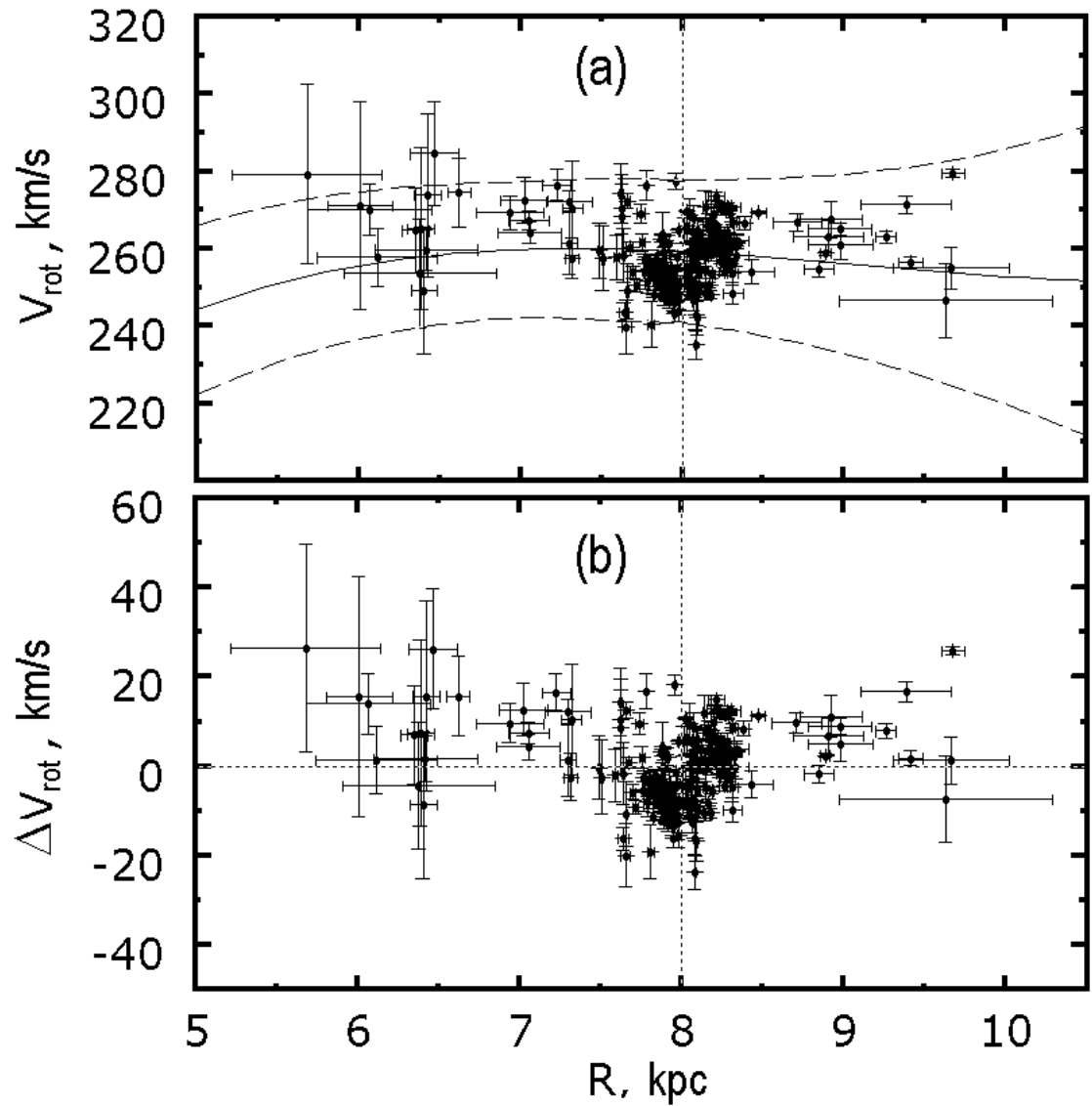}
  \caption{\small
(a)~Galactic rotation curve; the vertical dotted line marks the
position of the solar circle; the dashed lines indicate the
boundaries of the 1у confidence intervals. (b)~The residual
velocities obtained relative to the rotation curve shown above.
 }
 %%\label{f-XY}
 \end{center}}
 \end{figure}

\section{RESULTS AND DISCUSSION}
\subsection{Galactic Rotation Parameters}
First, we obtained the solution based on a sample of 97
spectroscopic binaries (the rejected systems are listed in Section
3.5) using only two equations, (2) and (3). This is because the
contribution from Eq. (4) to the general solution is negligible
when using distant stars. The velocity $w_\odot$ was assumed to be
known (because it is determined very poorly from distant stars)
and was taken to be $w_\odot = 7$~km s$^{-1}$. Such an approach
was used, for example, by Mishurov and Zenina (1999). The results
turned out to be the following:  $(u_\odot,v_\odot)=
(3.3,8.8)\pm(0.9,1.3)$~km s$^{-1}$ and
 \begin{equation}
 \renewcommand{\arraystretch}{1.2}
 \begin{array}{lll}
 \Omega_0      = 32.5 \pm1.1~\hbox{km s$^{-1}$ kpc$^{-1}$},  \\
 \Omega^{'}_0  = -4.41\pm0.19~\hbox{km s$^{-1}$ kpc$^{-2}$}, \\
 \Omega^{''}_0 =  0.97\pm0.37~\hbox{km s$^{-1}$ kpc$^{-3}$}, \\
            f_R=-11.5\pm1.9~\hbox{km s$^{-1}$},              \\
       f_\theta=  5.5\pm1.6~\hbox{km s$^{-1}$},              \\
              i= -6.4^\circ\pm0.6^\circ,             \\
 \chi_\odot=-117^\circ\pm6^\circ,
 \label{solution-1}
 \end{array}
 \end{equation}
The error per unit weight is $\sigma_0=8.5$~km s$^{-1}$. Based on
the derived pitch angle i and Eq. (9), we find
$\lambda=2.8\pm0.3$~kpc.

Since the solution using nearby Hipparcos B stars is of interest,
we obtained the solution using the entire sample (220 stars). All
three equations (2)--(4) were used in this approach; the velocity
$w_\odot$ was not fixed. The results turned out to be the
following: $(u_\odot,v_\odot,w_\odot)=
(3.4,8.9,7.8)\pm(0.8,1.1,0.2)$~km s$^{-1}$ and
 \begin{equation}
 \renewcommand{\arraystretch}{1.2}
 \begin{array}{lll}
 \Omega_0      = 32.4 \pm1.1~\hbox{km s$^{-1}$ kpc$^{-1}$},        \\
 \Omega^{'}_0  = -4.33\pm0.19~\hbox{km s$^{-1}$ kpc$^{-2}$}, \\
 \Omega^{''}_0 =  0.77\pm0.42~\hbox{km s$^{-1}$ kpc$^{-3}$}, \\
            f_R=-10.8\pm1.2~\hbox{km s$^{-1}$},             \\
       f_\theta=  7.9\pm1.3~\hbox{km s$^{-1}$},             \\
              i= -6.0^\circ\pm0.4^\circ,             \\
     \chi_\odot=-120^\circ\pm4^\circ,
 \label{solution-2}
 \end{array}
 \end{equation}
The error per unit weight is $\sigma_0=6.5$~km s$^{-1}$,
$\lambda=2.6\pm0.2$~kpc and the linear rotation velocity of the
Galaxy is $V_0=|R_0\Omega_0|=259\pm16$~km s$^{-1}$ (at
$R_0=8$~kpc).

From our comparison of results (19) and (20) we see that the most
distant stars have a decisive effect when finding the Galactic
rotation parameters. Using a large number of nearby stars with
highly accurate distances reduces only slightly the errors of the
parameters being determined. On the other hand, in solution (20)
the velocity $w_\odot$ is determined well and the tangential
perturbation amplitude $f_\theta$ increases.

The derived parameters of the Galactic rotation curve are in good
agreement with the results of analyzing young Galactic disk
objects rotating most rapidly around the center, such as OB
associations, $\Omega_0 =-31\pm1$~km s$^{-1}$ kpc$^{-1}$ (Mel\'nik
et al. 2001; Mel\'nik and Dambis 2009), blue supergiants,
 $\Omega_0=-29.6\pm1.6$~km s$^{-1}$ kpc$^{-1}$ and
 $\Omega'_0= 4.76\pm0.32$~km s$^{-1}$ kpc$^{-2}$ (Zabolotskikh et al. 2002),
or OB3 stars,
 $\Omega_0 = -31.5\pm0.9$~km s$^{-1}$ kpc$^{-1}$,
 $\Omega^{'}_0 = +4.49\pm0.12$~km s$^{-1}$ kpc$^{-2}$, and
 $\Omega^{''}_0 = -1.05\pm0.38$~km s$^{-1}$ kpc$^{-3}$ (Bobylev and Bajkova 2011).

As can be seen from Fig. 2, the patterns of change in the rotation
velocities of stars and hydrogen clouds are similar at $R>7$~kpc,
i.e., the phases of the velocity perturbations produced by the
spiral density wave are similar. The significant velocity
dispersion of OB stars at distances $R\approx6$~kpc is caused by
the proper motion errors.

  \begin{figure}[t]
 {\begin{center}
  \includegraphics[width=100mm]{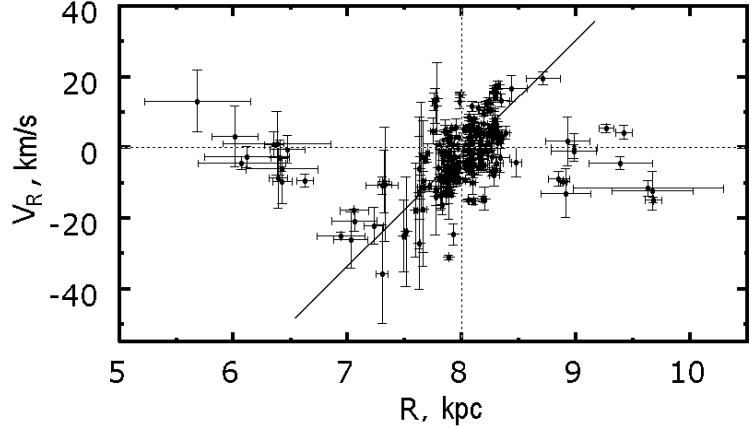}
  \caption{\small
Galactocentric radial velocities of the sample stars $V_R$ versus
distance $R$; the vertical line marks the position of the solar
circle.
 }
 %%\label{f-XY}
 \end{center}}
 \end{figure}

Figure 3a presents the rotation velocities of the stars from our
entire sample and the Galactic rotation curve constructed in
accordance with solution (20). The confidence intervals were
calculated by taking into account the uncertainty in $R_0.$ Figure
3b presents the residual velocities $\Delta V_{rot}$ calculated
using the derived Galactic rotation curve.

In Fig. 4, the Galactocentric radial velocities ($V_R$) of the
sample stars are plotted against distance $R$. The periodicity
related to the influence of the spiral density wave is clearly
seen. We plotted the line whose slope to the horizontal axis is
equal to the radial velocity gradient $dV_R/dR\approx40$~km
s$^{-1}$ kpc$^{-1}$ near $R=R_0$ (this gradient will be considered
in more detail in Section 3.3).

To construct the curves in Figs. 3 and 4, we used the stellar
velocities calculated with the following parameters of the local
standard of rest:
$(U_\odot,V_\odot,W_\odot)_{LSR}=(11.1,12.2,7.3)$ km s$^{-1}$
(Sch\"{o}nrich et al. 2010). It can be clearly seen from Fig. 4
that all distant stars are shifted downward by $\approx$6~km
s$^{-1}$ ($\Delta U_\odot$). The point is that so significant an
effect was detected neither in Bobylev and Bajkova (2011), where
the kinematics of OB3 stars was considered, nor in our papers
aimed at analyzing Galactic masers (Bobylev and Bajkova 2010;
Bajkova and Bobylev 2012).

  \begin{figure}[t]
 {\begin{center}
  \includegraphics[width=140mm]{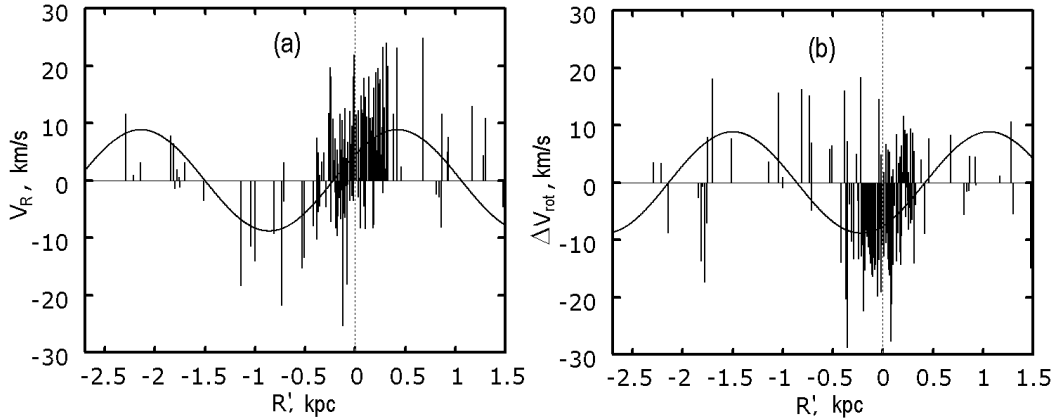}
  \caption{\small
(a) Galactocentric radial velocities $V_R$ for 220 stars versus
distance $R'$; (b) residual rotation velocities $\Delta V_{rot}$;
the velocities are given relative to the group velocity found.
 }
 %%\label{f-XY}
 \end{center}}
 \end{figure}

In Fig. 5, the radial, $V_R,$ and residual tangential, $\Delta
V_{rot}$, velocities for the stars of the entire sample are
plotted against distance $R'$, with these velocities being given
relative to the group velocity $(u_\odot,v_\odot,w_\odot)$ found
in solution (20). The power spectrum corresponding to this
approach is shown in Fig. 6.

Our comparison of solutions (19) and (20) and the results of our
spectral analysis of the residual velocities presented in Figs. 5
and 6 lead us to conclude that the derived spiral density wave
parameters agree well between themselves. Thus, the wavelength is
$\lambda=2.6$~kpc and the phase is $\chi_\odot=-120^\circ$.
However, these values still differ from $\lambda=2.3\pm0.2$~kpc
and $\chi_\odot=-91^\circ\pm4^\circ$ that were found in our
previous paper from OB3 stars using an independent distance scale
(Bobylev and Bajkova 2011), which will be discussed in
Section~3.4. As can be seen from Fig. 6, the spectral peak is well
extracted, although the spectrum has a large low-frequency
component.

  \begin{figure}[t]
 {\begin{center}
  \includegraphics[width=80mm]{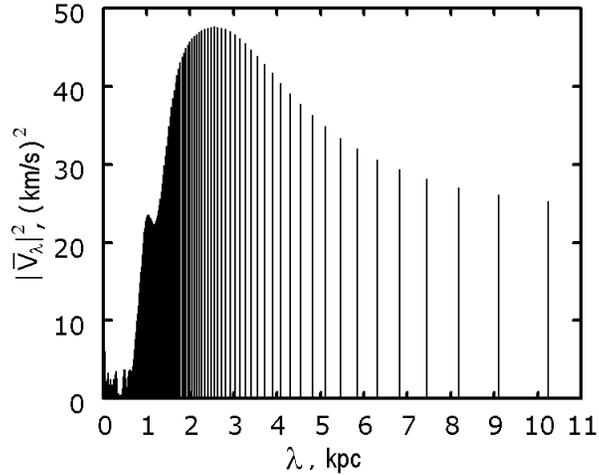}
  \caption{\small
Power spectrum of the velocities for the entire sample of stars.
 }
 %%\label{f-XY}
 \end{center}}
 \end{figure}

\subsection{Residual Velocities}
The stellar velocities corrected both for the Galactic
differential rotation and for the influence of the spiral density
wave are presented in Fig. 7 for distant stars ($r>0.6$~kpc) and
in Fig. 8 for nearby stars ($r\leq0.6$~kpc). The diagonal lines in
Figs. 7a and 8a indicate the orientation of the velocity ellipsoid
on the UV plane typical of the Gould Belt stars (Bobylev 2006). As
can be seen from Figs. 7d and 8d, applying a correction for the
influence of the spiral structure removes the elongation and makes
the distribution of stars nearly circular.

For example, for nearby stars before applying any corrections for
the influence of the density wave (this case corresponds to Figs.
8a--8c), the means are
 $({\overline U},{\overline V},{\overline W})=(-10.0,-14.7,-7.2)$~km s$^{-1}$
(as we see, the magnitudes of these velocities differ little from
the solar motion parameters derived by Sch\"{o}nrich et al.
(2010). In this case, the principal axes of the velocity ellipsoid
are
$(\sigma_U,\sigma_V,\sigma_W)=(7.3,5.7,3.8)\pm(0.5,0.6,0.4)$~km
s$^{-1}$; the orientation of the first axis ($l_1=308\pm7^\circ,
b_1=1\pm1^\circ$) is shown in Fig. 8a. Once the corrections for
the influence of the density wave have been applied (this case
corresponds to Figs. 8d.8f), the coordinates of the distribution
center and the velocity dispersions are
 $({\overline U},{\overline V},{\overline W})=(-4.5,-8.6,-7.2)$~km s$^{-1}$ and
 $(\sigma_U,\sigma_V,\sigma_W)=(6.1,5.8,3.7)\pm(0.6,0.5,0.4)$~km s$^{-1}$, and the
orientation of the velocity ellipsoid essentially coincides with
the directions of the $X,Y,Z$ coordinate axes.

\subsection{Kinematic Peculiarities of the Gould Belt}
When studying stars belonging to the Gould Belt, various authors
have long noticed such a kinematic effect as the expansion of this
entire star system with an expansion parameter $k\approx25$~km
s$^{-1}$ kpc$^{-1}$ (Bobylev 2006). A similar effect with an
expansion parameter $k\approx46$~km s$^{-1}$ kpc$^{-1}$ is also
detected in one of the components of the Gould Belt, the vast
Scorpius--Centaurus OB association (Bobylev and Bajkova 2007).
Torres et al. (2008) showed that the centers of all small
associations ($\beta$~Pic, Tuc-Hor, TW~Hya, and others) in the
neighborhood of Scorpius--Centaurus have an expansion parameter of
50~km s$^{-1}$ kpc$^{-1}$.

The radial velocity gradient $dVR/dR\approx40$~km s$^{-1}$
kpc$^{-1}$ indicated in Fig. 4 is very close to the above local
expansion parameters. However, remaining within the theory of the
influence of the spiral density wave, we must conclude that the
influence of the Galactic density wave should first be taken into
account to properly determine the intrinsic expansion parameters
of the Scorpius--Centaurus OB association and the entire Gould
Belt.

For example, the elongation of the velocity ellipsoid along the
diagonal in Fig. 8a is the most important indicator for the
presence of intrinsic expansion. contraction effects in the
system. However, as can be seen from Fig. 8d, such an elongation
is removed almost completely once the parameters of the spiral
density wave have been taken into account. This fact leaves fewer
chances for the implementation of a simple expansion model of the
Gould Belt as a whole. Studying this question is not the main task
of this paper; it will be considered in another paper.

Studying the influence of a large number of nearby Gould Belt
stars on our solutions (19) and (20) is also of interest. In
particular, the influence of the intrinsic kinematics of the Gould
Belt can be responsible for the discrepancy between the phase
angle $\chi_\odot=-120^\circ\pm4^\circ$ (solution (20)) and
$\chi_\odot=-91^\circ\pm4^\circ$ found from OB3 stars (at
$r>0.8$~kpc) using the ``calcium'' distance scale (Bobylev and
Bajkova 2011). To test this assumption, we obtained the solution
based on a sample free from Gould Belt stars.

  \begin{figure}[p]
 {\begin{center}
  \includegraphics[width=140mm]{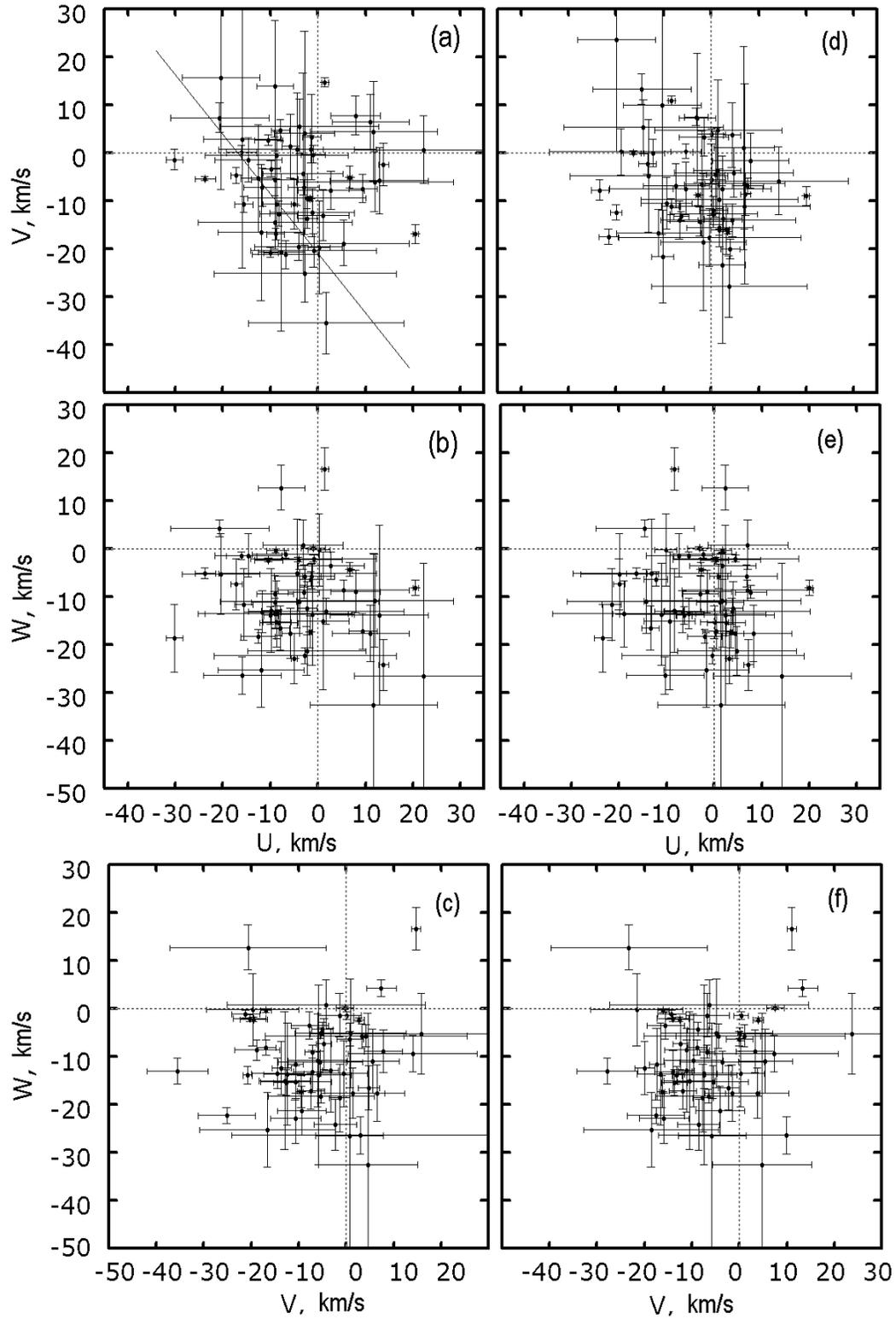}
  \caption{\small
Residual velocities $U,V,W$ of 58 distant stars relative to the
Sun: (a), (b), (c) corrected for the Galactic differential
rotation; (d), (e), (f) additionally corrected for the influence
of the Galactic spiral density wave.
 }
 %%\label{f-XY}
 \end{center}}
 \end{figure}

  \begin{figure}[p]
 {\begin{center}
  \includegraphics[width=140mm]{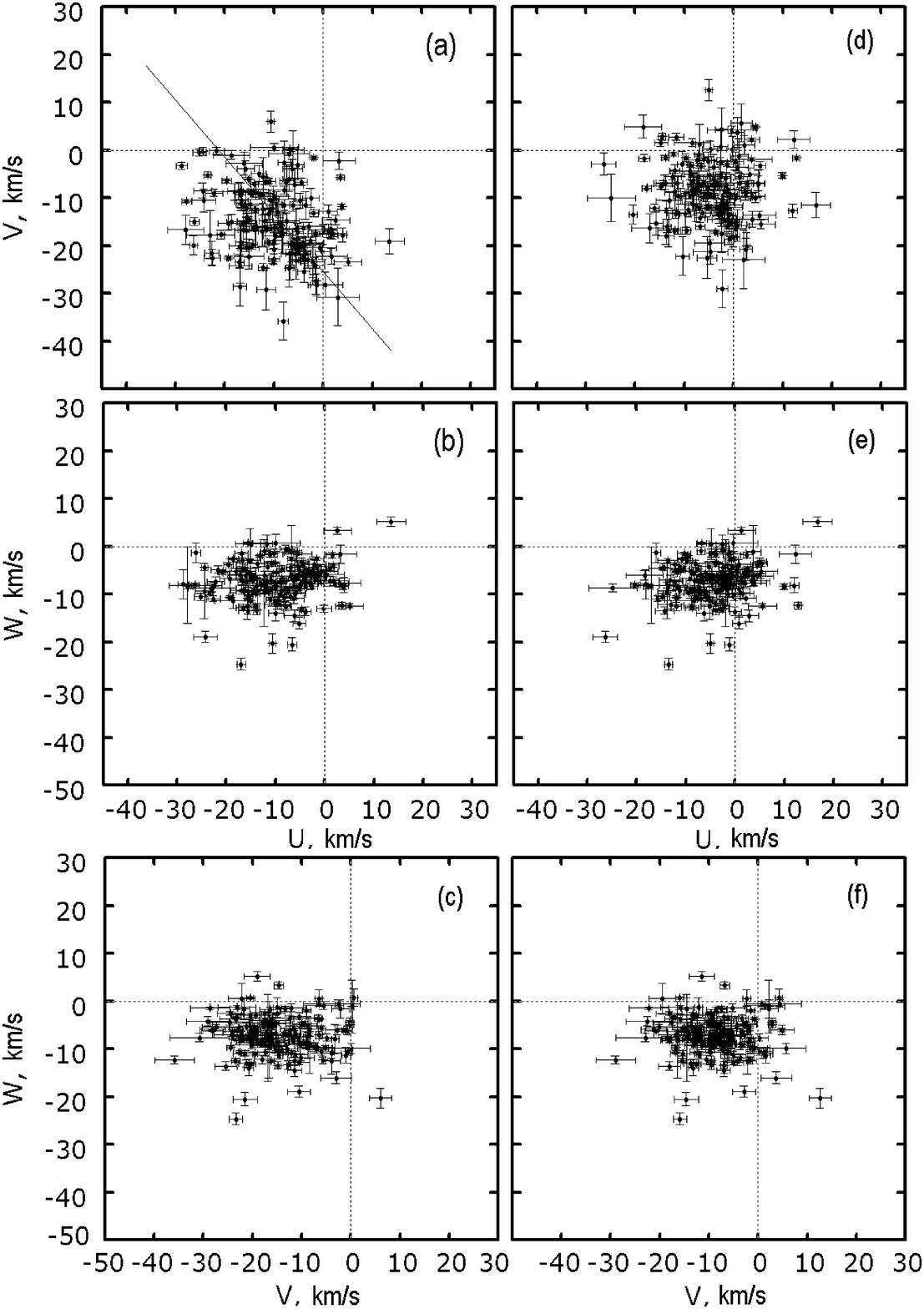}
  \caption{\small
Residual velocities $U,V,W$ of 162 nearby stars relative to the
Sun: (a), (b), (c) corrected for the Galactic differential
rotation; (d), (e), (f) additionally corrected for the influence
of the Galactic spiral density wave.
 }
 %%\label{f-XY}
 \end{center}}
 \end{figure}

\subsection{Galactic Kinematics from Distant Stars}
Based on the approach described when obtaining solution (19) and
using 58 distant stars located outside the circle with a radius of
0.6 kpc containing the Gould Belt stars (Fig.~1), we found the
following sought-for parameters: $(u_\odot,v_\odot)=
(2.2,8.4)\pm(0.9,1.1)$~km s$^{-1}$ and
 \begin{equation}
 \renewcommand{\arraystretch}{1.2}
 \begin{array}{lll}
 \Omega_0      = 31.9 \pm1.1~\hbox{km s$^{-1}$ kpc$^{-1}$},        \\
 \Omega^{'}_0  = -4.30\pm0.16~\hbox{km s$^{-1}$ kpc$^{-2}$}, \\
 \Omega^{''}_0 =  1.05\pm0.35~\hbox{km s$^{-1}$ kpc$^{-3}$}, \\
            f_R=-10.8\pm1.2~\hbox{km s$^{-1}$},      \\
       f_\theta=  2.9\pm2.1~\hbox{km s$^{-1}$},      \\
              i= -7.3^\circ\pm0.5^\circ,             \\
     \chi_\odot=-104^\circ\pm6^\circ.
 \label{solution-3}
 \end{array}
 \end{equation}
The error per unit weight is $\sigma_0 = 9.9$~km s$^{-1}$, the
wavelength is $\lambda = 3.2\pm0.3$~kpc, and $V_0=255\pm16$~km
s$^{-1}$. Figure~9 presents the results of our spectral analysis
of the radial velocities (our analysis of the residual tangential
velocities did not give a statistically significant perturbation
amplitude). The wavelength of the radial velocity perturbations is
$\lambda = 2.7$~kpc, which is slightly lower than that obtained by
solving the kinematic equations. The significance of the peak is
very high. Owing to the removal of nearby stars from the spectrum,
the low-frequency part decreased compared to the spectrum shown in
Fig. 6.

The velocity $V_0$ is of great importance for constructing an
appropriate Galactic rotation model and a model of the Galactic
potential. For example, the Galactic rotation curve constructed by
Sofue et al. (2009) with the adopted $R_0=8$~kpc and
 $V_0=200$~km s$^{-1}$ close to the parameters recommended by the IAU (1985) is well
known. However, a kinematic analysis of OB stars (Zabolotskikh et
al. 2002) or masers (Reid et al. 2009; McMillan and Binney 2010;
Bajkova and Bobylev 2012) shows that $V_0$ for the youngest disk
objects is slightly larger, being 240--260 km s$^{-1}$. Note the
paper by Irrgang et al. (2013), where three models of the Galactic
potential constructed using data on hydrogen clouds and masers
were proposed, with $V_0$ having been found to be close to 240 km
s$^{-1}$ (at $R_0\approx8.3$~kpc).

There are two parameters in solution (21), $\chi_\odot$ and
$f_\theta$, whose values differ significantly from those obtained
in solutions (19) and (20). Now, the amplitude ratio and the Sun's
phase are in agreement with the results of analyzing blue
supergiants,
 $f_R=-6.6\pm2.5$~km s$^{-1}$,
 $f_\theta= 0.4\pm2.3$~km s$^{-1}$, and
 $\chi_\odot=-97^\circ\pm18^\circ$ (Zabolotskikh et al. 2002), and in agreement with
 $f_R=-13\pm1$~km s$^{-1}$,
 $f_\theta=0\pm1$~km s$^{-1}$, and
 $\chi_\odot=-91^\circ\pm4^\circ$ found from a sample of 102 distant OB3 stars (Bobylev and Bajkova
2011).

The following conclusions can be reached: (1) Gould Belt stars
affect significantly the results of our analysis of the tangential
velocities---it is these stars that make a major contribution to
such a large $f_\theta=7.9\pm1.3$~km s$^{-1}$ in solution (20);
(2) including Gould Belt stars in the sample leads to a noticeable
shift of the Sun's phase in the density wave,
$\chi_\odot=-120^\circ\pm4^\circ$, compared to
$\chi_\odot=-104^\circ\pm6^\circ$ found from distant stars. Note
that there was no such problem when we determined the spiral
density wave parameters from Cepheids (Bobylev and Bajkova 2012),
because Cepheids are not associated kinematically with the Gould
Belt.

  \begin{figure}[t]
 {\begin{center}
  \includegraphics[width=140mm]{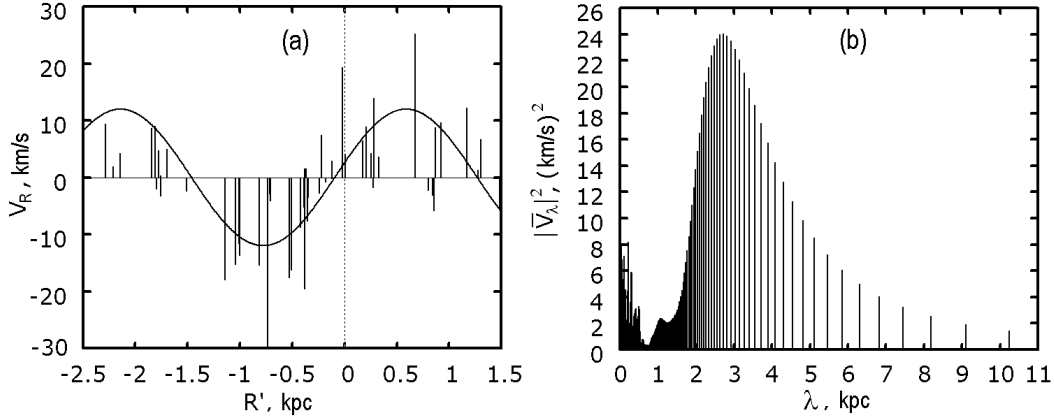}
  \caption{\small
(a) Galactocentric radial velocities $V_R$ of 58 distant stars
versus distance $R'$ and (b) their power spectrum.
 }
 %%\label{f-XY}
 \end{center}}
 \end{figure}

\subsection{Candidates for Runaway Stars}
The spiral density wave can produce a noticeable velocity
perturbation for some of the stars reaching 10--15 km s$^{-1}$,
depending on their positions in the spiral wave. Therefore, when
determining the status of a star as a runaway, we consider the
residual velocities calculated by taking into account the density
wave parameters found. We obtained the following results.

(1) According to the estimate by Torres et al. (2010), the
eclipsing spectroscopic binary DW~Car (B1V+B1V;
11M$_\odot$+11M$_\odot$) is located at a heliocentric distance of
$2840\pm150$~pc. It does not enter into the Hipparcos list of
stars. According to the UCAC4 data, its proper motion components
are
 $\mu_\alpha\cos\delta=-12.8\pm1.6$~mas yr$^{-1}$ and
 $\mu_\delta=-1.1\pm1.1$~mas yr$^{-1}$. Similar values were also obtained in
the TRC catalogue (Hog et al. 1998). The line-of-sight velocity is
$V_\gamma=-7.3\pm3.7$~km s$^{-1}$ (Southworth and Clausen 2007).
We found it to have a huge peculiar velocity at these parameters,
$|V_{pec}|=113\pm28$~km s$^{-1}$.

(2) The SB2 system DH Cep (O5.5V+O6.5V) has a significant residual
velocity, $|V_{pec}|=44\pm15$~km s$^{-1}$. We used a distance
estimate of $2767\pm450$~pc for it (Hilditch et al. 1996).
However, it is not marked as a runaway in the GOS2 catalogue
(Maiz-Apella\'niz et al. 2004).

(3) The distant SB systemHD1383 (B0.5I+B0.5I) located at a
distance of $2860\pm570$~pc (Gudennavar et al. 2012) has
$|V_{pec}|=40\pm14$~km s$^{-1}$.

(4) HD 150136 (HR 6187; O3V+O5.5+O6.5V) located at a distance of
$1320\pm120$~pc recently classified by Mahy et al. (2012) as a
spectroscopic triple system has $|V_{pec}|=36\pm10$~km s$^{-1}$.

(5) The binary system V~1034 Sco (O9V+B1.5V;
17M$_\odot$+9.6M$_\odot$) from the list by Torres et al. (2010),
$r=1640\pm89$~pc, does not enter into the Hipparcos list of stars.
It does not have a large peculiar velocity ($|V_{pec}|=28\pm17$~km
s$^{-1}$), but it is always rejected according to the $3\sigma$
criterion. The proper motion components (UCAC4) for it are
probably not quite reliable.

(6) Finally, the star HIP 99303 (V1624 Cyg, B2.5V),
$r=318\pm23$~pc, is rejected according to the $3\sigma$ criterion,
although its peculiar velocity is small, $|V_{pec}|=28\pm2$~km
s$^{-1}$.

As can be seen from this list, two stars, DW Car and the SB2
system DH Cep, can be attributed with a high degree of reliability
to runaway stars, because their peculiar velocities exceed
40~km~s$^{-1}$. The star DW~Car whose peculiar velocity exceeds
100~km~s$^{-1}$ especially stands out.

We excluded the listed stars when solving Eqs. (2)--(4). However,
as our analysis that is not presented here showed, the
line-of-sight velocities of these stars are quite acceptable and
can be used in Eqs. (2)--(4). They are not rejected according to
the $3\sigma$ criterion and do not spoil the solutions.

\section{DESCRIPTION OF THE DATABASE}
The data format is described in Table 3. Note that two values are
given for the line-of-sight velocity error, $\varepsilon_{V_r}$
and $\sigma_{V_r}$. The quantity $\varepsilon_{V_r}$ is the formal
random error in the systemic velocity $V_\gamma$ when interpreting
the line-of-sight velocity curves; its value can be very low,
$<0.05$~km~s$^{-1}$, for a number of stars (in such cases, they
were rounded off to 0.1~km~s$^{-1}$). The quantity $\sigma_{V_r}$
characterizes the quality of the observed line-of-sight velocities
as a whole---we used this quantity to assign the individual
weights of the stars when solving the system of equations
(2)--(4). As $\sigma_{V_r}$ we took either the dispersion of the
residuals that the authors usually specify (r.m.s) and, in the
case of its absence, the errors in the velocity amplitude for the
first component (K1) of a spectroscopic binary system.

The peculiar space velocities of the stars were calculated by
taking into account the Galactic differential rotation effects and
the influence of the spiral density wave with the parameters found
in solution~(20). Information about the bibliography is given in
Table~1.

%%%%%%%%%%%%%%%%%%%%%%%%%%%%%%%%%%%%%%%%%%%%%%%%%%%%%%%%%%%%%%%%%%%%%%%%%%%%%%%%%%%%%%%%%%%%%%%%%%%%%%
\begin{table}[t]                                                %% T3
 \label{t:03}
\caption[]{\small\baselineskip=1.0ex\protect
 Format of the database}
\begin{center}
\begin{tabular}{|l|l|}\hline

Alternative star name                                       & A12   \\
HD/CD/BD number                                             & A12   \\
Hipparcos number                                            & A10   \\
Right ascension in degrees                                  & F11.6 \\
Declination in degrees                                      & F11.6 \\
Heliocentric distance in pc                                 & F6.1  \\
Proper motion in right ascension in mas yr$^{-1}$           & F6.2  \\
Proper motion in declination in mas yr$^{-1}$                  & F6.2  \\
Line-of-sight velocity in km s$^{-1}$                       & F6.1  \\
Heliocentric distance error in pc                           & F5.1  \\
Error of proper motion in right ascension in mas yr$^{-1}$  & F4.2  \\
Error of proper motion in declination in mas yr$^{-1}$      & F4.2  \\
Line-of-sight velocity error in km s$^{-1}$                 & F4.1  \\
Line-of-sight velocity dispersion in km s$^{-1}$            & F4.1  \\
Spectral type of first component                            & A10   \\
Spectral type of second component                           & A8    \\
Spectral type of third component                            & A5    \\
Peculiar space velocity in km s$^{-1}$                      & F5.1  \\
Error of peculiar space velocity in km s$^{-1}$             & F4.1  \\
 Reference information                                      & A14\\\hline
\end{tabular}
\end{center}
\end{table}
%%%%%%%%%%%%%%%%%%%%%%%%%%%%%%%%%%%%%%%%%%%%%%%%%%%%%%%%%%%%%%%%%%%%%%%%%%%%%%%%%%

\section*{CONCLUSIONS}
Based on published sources, we created a kinematic database on
massive ($>10M_\odot$) young Galactic star systems located within
$\leq3$~kpc of the Sun. For most of the stars within
$\approx$600~pc of the Sun, the distance errors do not exceed
10\%. We included all massive OB stars with known estimates of
their orbital (dynamical) parallaxes; the errors in the dynamical
distances for seven such systems are smaller than those in the
trigonometric ones, with the errors for two of them being equal.
Spectrophotometric distance estimates were used for more distant
stars, but there are also calibration spectroscopic binary systems
with very reliable distances among these objects. The main
advantage of the proposed list of spectroscopic binary systems is
that the latest information about the measured line-of-sight
velocities and proper motions is reflected for them.

Based on the entire sample of 220 stars, we found the circular
rotation velocity of the solar neighborhood around the Galactic
center, $V_0=259\pm16$~km s$^{-1}$ (at $R_0=8$~kpc), and the
following spiral density wave parameters: the amplitudes of the
radial and azimuthal velocity perturbations $f_R=-10.8\pm1.2$~km
s$^{-1}$ and $f_\theta= 7.9\pm1.3$~km s$^{-1}$, respectively; the
pitch angle for a two-armed spiral pattern
$i=-6.0^\circ\pm0.4^\circ$ ($\lambda=2.6\pm0.2$~kpc); and the
radial phase of the Sun in the spiral density wave
$\chi_\odot=-120^\circ\pm4^\circ$. The residual velocity
dispersion for the stars of the entire sample obtained by taking
into account the Galactic differential rotation effects and the
influence of the spiral density wave is $\sigma\approx7$~km
s$^{-1}$. This suggests that the stars are young and that their
kinematic characteristics are close to those of hydrogen clouds,
for which the residual velocity dispersion is 5.6 km s$^{-1}$.

Based on a sample of 58 distant ($r>0.6$~kpc) stars, we found the
following parameters:
 $V_0=255\pm16$~km s$^{-1}$,
 $f_R=-10.8\pm1.2$~km s$^{-1}$,
 $f_\theta= 2.9\pm2.1$~km s$^{-1}$,
 $i=-7.3^\circ\pm0.5^\circ$, and
 $\chi_\odot=-104^\circ\pm6^\circ$. These are in good agreement with the results of
analyzing other samples of OB stars by other authors. These
parameters describe the properties of the Galactic grand-design
spiral structure in the neighborhood under consideration.

Such peculiarities of the Gould Belt as the local expansion of the
system, the velocity ellipsoid vertex deviation, and the
significant additional rotation can be explained in terms of the
density wave theory. All these effects decrease significantly once
the influence of the spiral density wave on the velocities of
these stars has been taken into account. Our study of the
contribution from Gould Belt stars to the determination of the
spiral density wave parameters showed that (1) these stars affect
significantly the results of analyzing the tangential
velocities---they make a major contribution to the large value of
$f_\theta=7.9\pm1.3~\hbox{км/c}$~km s$^{-1}$; (2) including Gould
Belt stars in the sample leads to a noticeable shift of the Sun's
phase in the density wave, $\Delta\chi_\odot=30^\circ$, compared
to the value found from distant ($r>0.6$~kpc) stars. Among the
list of calibration spectroscopic binary systems by Torres et al.
(2010), the runaway star DW Car with a significant peculiar
velocity, $|V_{pec}|=113\pm28$~km s$^{-1}$, has probably been
revealed for the first time. The SB2 system DH~Cep also has an
appreciable peculiar velocity, $|V_{pec}|=44\pm15$~km s$^{-1}$.

\section*{ACKNOWLEDGMENTS}
We are grateful to the referee for helpful remarks that
contributed to a improvement of the paper. This work was supported
by the ``Nonstationary Phenomena in Objects of the Universe''
Program of the Presidium of the Russian Academy of Sciences and
the ``Multiwavelength Astrophysical Research'' grant no.
NSh--16245.2012.2 from the President of the Russian Federation. In
our work, we widely used the SIMBAD astronomical database.

\section*{REFERENCES}
{\small

\quad~~1. E. Anderson and S. Francis, Astron. Lett. 38, 331
(2012).

2. M. Ausseloos, C. Aerts, K. Lefever, et al., Astron. Astrophys.
455, 259 (2006).

3. V.C. Avedisova, Astron. Rep. 49, 435 (2005).

4. A.T. Bajkova and V.V. Bobylev, Astron. Lett. 38, 549 (2012).

5. H. Bakis, V. Bakis, O. Demircan, et al., Mon. Not. R. Astron.
Soc. 385, 381 (2008).

6. V. Bakis, I. Bulut, S. Bilir, et al., Publ. Astron. Soc. Jpn.
62, 1291 (2010).

7. S.A. Bell, D. Kilkenny, and G.J. Malcolm, Mon. Not. R. Astron.
Soc. 226, 879 (1987).

8. V.V. Bobylev, Astron. Lett. 32, 816 (2006).

9. V.V. Bobylev and A.T. Bajkova, Astron. Lett. 33, 571 (2007).

10. V.V. Bobylev and A.T. Bajkova, Mon. Not. R. Astron. Soc. 408,
1788 (2010).

11. V.V. Bobylev and A.T. Bajkova, Astron. Lett. 37, 526 (2011).

12. V.V. Bobylev and A.T. Bajkova, Astron. Lett. 38, 638 (2012).

13. V.V. Bobylev, A.T. Bajkova, and A. S. Stepanishchev, Astron.
Lett. 34, 515 (2008).

14. R.L. Branham, Mon. Not. R. Astron. Soc. 370, 1393 (2006).

15. J.A. Caballero, Mon. Not. R. Astron. Soc. 383, 750 (2008).

16. D.P. Clemens, Astrophys. J. 295, 422 (1985).

17. A. Coleiro and S. Chaty, arXiv:1212.5460 (2012).

18. Z. Cvetkovi\'c , I. Vince, and S. Ninkovi\'c, New Astron. 15,
302 (2010).

19. J.A. Docobo and M. Andrade, Astrophys. J. 652, 681 (2006).

20. S.A. Dzib, L.F. Rodriguez, L.Loinard, et al., arXiv:1212.1498
(2012).

21. T. Foster and B. Cooper, astro-ph: 1009.3220 (2010).

22. G.A. Gontcharov, Astron. Lett. 32, 759 (2006).

23. S.B. Gudennavar, S.G. Bubbly, K. Preethi et al., Astrophys. J.
Suppl. Ser. 199, 8 (2012).

24. T.J. Harries, R.W. Hilditch, and G. Hill, Mon. Not. R. Astron.
Soc. 285, 277 (1997).

25. R.W. Hilditch, T.J. Harries, and S.A. Bell, Astron. Astrophys.
314, 165 (1996).

26. E. Hog, A. Kuzmin, U. Bastian, et al., Astron. Astrophys. 335,
L65 (1998).

27. E. Hog, C. Fabricius, V.V. Makarov, et al., Astron. Astrophys.
Lett. 355, L27 (2000).

28. M.M. Hohle, R. Neuh\aa user, and B.F. Schutz, Astron. Nachr.
331, 349 (2010).

29. D.E. Holmgren, C.D. Scarfe, G. Hill, et al., Astron.
Astrophys. 231, 89 (1990).

30. K. van der Hucht, New Astron. Rev. 45, 135 (2001).

31. A. Irrgang, B. Wilcox, E. Tucker, et al., Astron. Astrophys.
549, 137 (2013).

32. M. Kennedy, S.M. Dougherty, A. Fink, et al., Astrophys. J.
709, 632 (2010).

33. N.V. Kharchenko, A.E. Piskunov, S. Roeser, et al., Astron.
Astrophys. 438, 1163 (2005).

34. N.V. Kharchenko, R.-D. Scholz, A.E. Piskunov, et al., Astron.
Nachr. 328, 889 (2007).

35. M.K. Kim, T. Hirota, M. Honma, et al., Publ. Astron. Soc. Jpn.
60, 991 (2008)

36. S. Kraus, G. Weigelt, Yu.Yu. Balega, et al., Astron.
Astrophys. 497, 195 (2009).

37. C.H.S. Lacy, A. Claret, and J.A. Sabby, Astron. J. 128, 1840
(2004).

38. F. van Leeuwen, Astron. Astrophys. 474, 653 (2007).

39. E.S. Levine, C. Heiles, and L. Blitz, Astrophys. J. 679, 1288
(2008).

40. C.C. Lin and F.H. Shu, Astrophys. J. 140, 646 (1964).

41. K.P. Lindroos, Astron. Astrophys. Suppl. Ser. 60, 183 (1985).

42. R. Lorenz, P.Mayer, and H. Drechsel, Astron. Astrophys. 332,
909 (1998).

43. A. Lutovinov, M. Revnivtsev, M. Gilfanov, et al., Astron.
Astrophys. 444, 821 (2005).

44. L. Mahy, Y. Naz\'e, G. Rauw, et al., Astron. Astrophys. 502,
937 (2009).

45. L. Mahy, G. Rauw, F. Martins, et al., Astrophys. J. 708, 1537
(2010).

46. L. Mahy, E. Gosset, H. Sana, et al., Astron. Astrophys. 540,
97 (2012).

47. L. Mahy, G. Rauw, and M. De Becker, astro-ph: 1301.0500
(2013).

48. J. Maiz-Apell\'aniz, N.R. Walborn, H.A. Galue, and L.H. Wei,
Astrophys. J. Suppl. Ser. 151, 103 (2004).

49. J. Maiz-Apell\'aniz, E.J. Alfaro, and A. Sota, astro-ph:
0804.2553 (2008).

50. W.L.F. Marcolino, F.X. de Ara\'ujo, S. Lorenz- Martins, et
al., Astron. J. 133, 489 (2007).

51. N.M. McClure-Griffiths, and J.M. Dickey, Astrophys. J. 671,
427 (2007).

52. P.J. McMillan and J.J. Binney, Mon. Not. R. Astron. Soc. 402,
934 (2010).

53. M.V. McSwain, T.S. Boyajian, E.D. Grundstrom, et al.,
Astrophys. J. 655, 473 (2007).

54. A. Megier, A. Strobel, G.A. Galazutdinov, et al., Astron.
Astrophys. 507, 833 (2009).

55. A.M. Mel’nik and A.K. Dambis, Mon. Not. R. Astron. Soc. 400,
518 (2009).

56. A.M. Mel’nik, A.K. Dambis, and A.S. Rastorguev, Astron. Lett.
27, 521 (2001).

57. Yu.N. Mishurov and I.A. Zenina, Astron. Astrophys. 341, 81
(1999).

58. A.E.J. Moffat, S.V. Marchenko, W. Seggewiss, et al., Astron.
Astrophys. 331, 949 (1998).

59. J.R. North, P.G. Tuthill, W.J. Tango, and J. Davis, Mon. Not.
R. Astron. Soc. 377, 415 (2007a).

60. J.R. North, J. Davis, P.G. Tuthill, et al., Mon. Not. R.
Astron. Soc. 380, 1276 (2007b).

61. P. Patriarchi, L. Morbidelli, and M. Perinotto, Astron.
Astrophys. 410, 905 (2003).

62. L.R. Penny, D.R. Gies, J.H. Wise, et al., Astrophys. J. 575,
1050 (2002).

63. M.E. Popova and A.V. Loktin, Astron. Lett. 31, 663 (2005).

64. D. Pourbaix, A.A. Tokovinin, A.H. Batten, et al., Astron.
Astrophys. 424, 727 (2004).

65. A.S. Rastorguev, E.V. Glushkova, A.K. Dambis, and M.V.
Zabolotskikh, Astron. Lett. 25, 595 (1999).

66. G. Rauw, J.-M. Vreux, and B. Bohannan, Astrophys. J. 517, 416
(1999).

67. G. Rauw, H. Sana, M. Spano, et al., Astron. Astrophys. 542, 95
(2012).

68. M.J. Reid, K.M. Menten, X.W. Zheng, et al., Astrophys. J. 700,
137 (2009).

69. N.D. Richardson, D.R. Gies, and S.J. Williams, Astron. J. 142,
201 (2011).

70. S. Roeser, M. Demleitner, and E. Schilbach, Astron. J. 139,
2440 (2010).

71. H. Sana, E. Antokhina, P. Royer, et al., Astron. Astrophys.
441, 213 (2005).

72. H. Sana, E. Gosset, and G. Rauw, Mon. Not. R. Astron. Soc.
371, 67 (2006).

73. H. Sana, G. Rauw, and E. Gosset, Astron. Astrophys. 370, 121
(2001).

74. H. Sana,G. Rauw, and E. Gosset,Astrophys. J. 659, 1582 (2007).

75. H. Sana, Y. Naz\'e, B. O’Donnell, et al., New Astron. 13, 202
(2008).

76. R. Sch\"{o}nrich, J. Binney, and W. Dehnen, Mon. Not. R.
Astron. Soc. 403, 1829 (2010).

77. Y. Sofue, M. Honma, and T. Omodaka, Publ. Astron. Soc. Jpn.
61, 227 (2009).

78. J. Southworth and J.V. Clausen, Astron. Astrophys. 461, 1077
(2007).

79. W.J. Tango, J. Davis, M.J. Ireland, et al., Mon. Not. R.
Astron. Soc. 370, 884 (2006).

80. D. Terrell, U. Munari, and A. Siviero, Mon. Not. R. Astron.
Soc. 374, 530 (2007).

81. N. Tetzlaff, R. Neuh\aa user, and M.M. Hohle, Mon. Not. R.
Astron. Soc. 410, 190 (2011).

82. The HIPPARCOS and Tycho Catalogues, ESA SP--1200 (1997).

83. C.A.O. Torres, G.R. Quast, C.H.F. Melo, et al., Handbook of
Star Forming Regions, Vol. II: The Southern Sky, ASP Monograph
Publ., Ed. by Bo Reipurth, arXiv:0808.3362v1 (2008).

84. G. Torres, J. Andersen, and A. Gime\'nez, Astron. Astrophys.
Rev. 18, 67 (2010).

85. C. Tycner, A. Ames, R.T. Zavala, et al., Astrophys. J. Lett.
729, L5 (2011).

86. K. Uytterhoeven, B. Willems, K. Lefever, et al., Astron.
Astrophys. 427, 581 (2004).

87. A.A. Wachmann, D.M. Popper, and J.V. Clausen, Astron.
Astrophys. 162, 62 (1986).

88. S.J. Williams, D.R. Gies, and T.C. Hillwig, Astron. J. 145, 29
(2013).

89. M.V. Zabolotskikh, A.S. Rastorguev, and A.K. Dambis, Astron.
Lett. 28, 454 (2002).

90. N. Zacharias, C.T. Finch, T.M. Girard, et al., I/322
Catalogue, Strasbourg Data Base (2012).

91. J. Zi\'o\l kowski, Mon. Not. R. Astron. Soc. 358, 851 (2005).

}

\end{document}